\documentclass[aps,pra,11pt,tightenlines]{revtex4-1}
\usepackage{mathtools}
\usepackage{units}
\usepackage{amssymb}
\usepackage{amsmath}
\usepackage{amsthm}
\usepackage{graphicx} 
\usepackage{bm}
\usepackage{braket}
\usepackage[usenames,dvipsnames,svgnames,table]{xcolor}
\usepackage{marginnote}
\newcommand{\pr}{\nabla_{\vec{r}}}
\newcommand{\pR}{\nabla_{\vec{R}}}
\newcommand{\pt}{\partial_t}

\newcommand{\px}{\nabla_{\vec{x}}}

\newcommand{\pop}{\vec{\hat{p}}}
\newcommand{\Pop}{\vec{\hat{P}}}
\newcommand{\popc}{\vec{\hat{p}}_{\rm c}}
\newcommand{\Popc}{\vec{\hat{P}}_{\rm c}}

\renewcommand{\vec}[1]{\bm{#1}}

\begin{document}
  
  \author{Axel Schild}
  \title{Time in quantum mechanics: \\ A fresh look on quantum hydrodynamics and 
         quantum trajectories}
  \affiliation{ETH Z\"urich,  Laboratorium f\"ur Physikalische Chemie,  8093 Z\"urich, Switzerland}
  
  \begin{abstract}
    Quantum hydrodynamics is a formulation of quantum mechanics based on the 
    probability density and flux (current) density of a quantum system.
    It can be used to define trajectories which allow for a particle-based 
    interpretation of quantum mechanics, commonly known as Bohmian mechanics.
    However, quantum hydrodynamics rests on the usual time-dependent 
    formulation of quantum mechanics where time appears as a parameter.
    This parameter describes the correlation of the state of the quantum system
    with an external system -- a clock -- which behaves according to classical 
    mechanics.
    With the Exact Factorization of a quantum system into a marginal and a 
    conditional system, quantum mechanics and hence quantum hydrodynamics can 
    be generalized for quantum clocks.
    In this article, the theory is developed and it is shown that trajectories 
    for the quantum system can still be defined, and that these trajectories 
    depend conditionally on the trajectory of the clock.
    Such trajectories are not only interesting from a fundamental point of view, 
    but they can also find practical applications whenever a dynamics relative 
    to an external time parameter is composed of ``fast'' and ``slow'' degrees 
    of freedom and the interest is in the fast ones, while quantum effects of 
    the slow ones (like a branching of the wavepacket) cannot be neglected.
    As an illustration, time- and clock-dependent trajectories are calculated 
    for a model system of a non-adiabatic dynamics, where an electron is the 
    quantum system, a nucleus is the quantum clock, and an external time 
    parameter is provided, e.g.\ via an interaction with a laser field that is 
    not treated explicitly.
    
  \end{abstract}
  
  \maketitle
  
  Although being developed for ca.\ 100 years, the meaning of quantum mechanics
  is still a topic of active discussion.
  Next to the open question of merging quantum mechanics with general relativity,\cite{isham1993,anderson2017,bose2017,carney2019} 
  the development of first usable quantum computers \cite{preskill2018} has 
  recently sparked interest in the fundamentals of the theory and questions 
  like its 
  consistent use attracted some attention.\cite{frauchiger2018,lazarovici2019}
  During the years, several formulations and interpretations of quantum mechanics have been proposed.\cite{styer2002}
  While they all may reproduce the known experimental results, some of them 
  are useful beyond philosophical questions while others have so far been of 
  little use other than for entertaining debates.
  Two closely related formulations of the former type are quantum hydrodynamics 
  and Bohmian mechanics,\cite{madelung1926,madelung1927,broglie1927,bohm1952,bohm1952_2} 
  which both suggest strategies for calculating properties of larger quantum 
  systems.
  Here, these two formulations are re-considered in the light of the emergence 
  of the concept of time in quantum mechanics.
  
  The standard approach to non-relativistic quantum mechanics is based on the time-dependent 
  Schr\"odinger equation (TDSE) that describes the state of a quantum system 
  by means of a wavefunction.
  It is the central quantity of the theory and contains all relevant information
  that is needed to compute observables of the quantum system.
  The wavefunction changes as time passes and, in this way, the state evolves.
  However, there are at least two problems with the theory of quantum mechanics:
  The first problem is the ``measurement problem'', as the act of measuring the 
  quantum system is usually treated as additional postulate.\cite{bell1990,hollowood2016,landsman2017}
  An interaction with an external system, the measurement apparatus, leads to a 
  ``collapse'' of the wavefunction into the state corresponding to the 
  measurement outcome.
  The second problem is the ``time problem'', as time in the TDSE is not a 
  quantum-mechanical observable but a parameter which leads to conceptual 
  challenges like the definition of a tunneling time \cite{steinberg1995,landsman2015,zimmermann2016,sokolovski2018} or the 
  unification of quantum mechanics with general relativity.\cite{isham1993,anderson2017}
  Both the measurement problem and the time problem are rooted in an 
  inconsistent treatment of the quantum system and its (external) environment.
  
  Two useful alternatives to a wavefunction-based picture can be 
  obtained by writing the complex-valued wavefunction of a quantum system in 
  its polar form.
  Then, equations of motion follow for a probability density and probability 
  flux (current) density.
  These equations are similar in appearance to the equations of motion of fluid 
  dynamics, hence they are known as quantum hydrodynamics \cite{madelung1926,madelung1927}
  and have found several applications, e.g.\ in the study of Bose-Einstein 
  condensates\cite{tsubota2013} or in plasma physics.\cite{shukla2011,moldabekov2018} 
  Recently, the extension of quantum hydrodynamics to many-particle systems has 
  also been studied in detail.\cite{renziehausen2018}
  
  Closely related to quantum hydrodynamics is the De Broglie-Bohm interpretation
  of quantum mechanics, also known as Bohmian mechanics, where particles are 
  assumed to follow trajectories with a velocity field that is determined by the 
  wavefunction.\cite{broglie1927,bohm1952,bohm1952_2}
  Bohmian mechanics is conceptually attractive for several reasons, e.g.\ 
  because some consider it not to have the measurement problem.\cite{durr1992,durr2004_2,norsen2016}
  For practical applications, Bohmian mechanics can be viewed as 
  trajectory-based quantum hydrodynamics, and in the last years a number of 
  studies have appeared that aim at using these quantum trajectories to develop 
  simulation methods, in particular for molecular dynamics \cite{wyatt2001,lopreore2002,wyatt2005,rassolov2005,curchod2013,agostini2018},
  also by extending it to complex-valued trajectories \cite{goldfarb2006,zamstein2012,zamstein2012_2,koch2017}
  or by using the ``conditional wavefunction'' approach \cite{oriols2007,benseny2014,albareda2014,albareda2016}.
  
  Both quantum hydrodynamics and Bohmian mechanics are derived from a 
  time-dependent description of quantum mechanics. 
  Hence, although Bohmian mechanics might not have the measurement problem, 
  it does have the time problem.
  While the time problem is often not a problem in practice, it is assumed to 
  be major theoretical problem for the unification of quantum mechanics with 
  the general theory of relativity.\cite{isham1993,anderson2017}
  A few solutions have been proposed, e.g.\ the Page-Wootters approach,\cite{page1983,wootters1984} 
  which gained some interest recently,\cite{moreva2014,moreva2015,giovannetti2015,moreva2017,marletto2017,bryan2018,smith2019} 
  or an approach based on the 
  Born-Oppenheimer approximation, which was applied to gravity and matter \cite{banks1985,brout1987,englert1989} 
  and which has also been considered by Briggs and Rost a few years ago.\cite{briggs2000}
  Additionally, time in general relativity has been reconsidered and there are 
  promising developments like shape dynamics \cite{barbour2014a,barbour2014b} 
  that illuminate the meaning of time as a correlation.
  All these show that the concept of an external time parameter needs to be 
  replaced with the explicit consideration some part of the system as the 
  ``clock'' that is used to define time.
  
  Here, we use an extension of the developments made by Briggs and Rost \cite{briggs2000}
  and by Briggs and coworkers \cite{braun2004,briggs2014,briggs2015} to derive 
  the time parameter in quantum mechanics. 
  With the help of this approach, it can be shown that ``time'' appearing in 
  the TDSE has a similar status like the measurement apparatus in standard 
  quantum mechanics:
  It is not truly part of the theory.
  Instead, it is a Newtonian time parameter $t$ that describes the correlation 
  of the state of the quantum system relative to an external system, the clock, 
  which provides $t$.
  To be able to do that, the clock needs to behave according to classical 
  mechanics so that e.g.\ its position and velocity are known and can be used 
  to define $t$.
  
  \begin{figure}[htbp]
    \centering
    \includegraphics[width=0.4\textwidth]{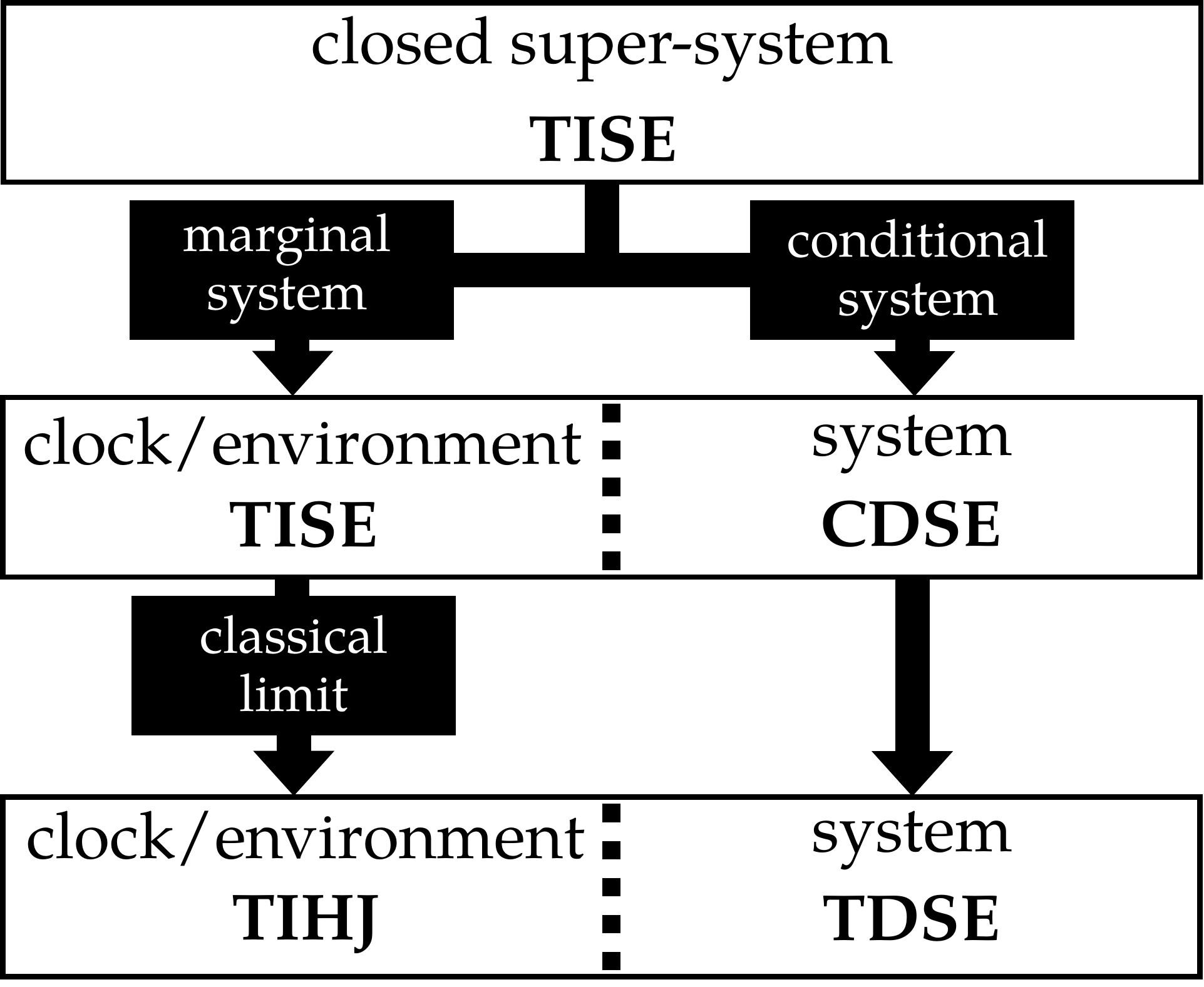}
    \caption{Schematic representation of the steps needed to derive time in 
             quantum mechanics: By separating a closed super-system described 
             by a time-independent Schr\"odinger equation (TISE) into a marginal 
             part (clock) and a conditional part (system), it is found that the 
             clock obeys an effective TISE and the system obeys a clock-dependent 
             Schr\"odinger equation (CDSE). If a classical limit is taken for 
             the clock, it obeys a time-independent Hamilton-Jacobi equation
             (TIHJ); the CDSE for the system system then becomes a 
             time-dependent Schr\"odinger equation (TDSE) for a time-parameter 
             defined via the classical clock.}
    \label{fig:scheme}
  \end{figure}
  
  Explicitly, the time-dependence of the TDSE can be derived from a
  time-independent description of a super-system which is being partitioned 
  into the actual system of interest and the clock,\cite{briggs2000,briggs2015,schild2018}
  as indicated in Figure \ref{fig:scheme}.
  The super-system is assumed to be closed and hence is described by a 
  time-independent Schr\"odinger equation (TISE).
  To introduce the concept of time in this time-independent description, a
  part of the super-system is considered to be a marginal part, i.e., its state 
  is being described after averaging over the rest of the super-system.
  This part is the clock, but it may also be viewed as an environment.
  The state of the rest of the super-system, which is the actual system of 
  interest, is then conditionally depended on the state of the clock.
  If a classical limit for the clock is taken, its equation of motion becomes a 
  classical time-independent Hamilton-Jacobi equation.
  A time parameter can then be defined along the classical trajectory of the 
  clock, and the equation of motion for the system of interest becomes a TDSE.\cite{briggs2015}
  It follows that, because time in the TDSE is defined via a classical clock, 
  the TDSE is actually a semi-classical equation.
  For the same reason, also quantum hydrodynamics as well as Bohmian mechanics are semi-classical 
  theories.
  
  The formalism for separating the quantum super-system into a marginal part 
  and a conditional part is called the Exact Factorization,\cite{abedi2010,abedi2012}
  which can be viewed as a rigorous extension of ideas from the Born-Oppenheimer
  approximation.\cite{born1927}
  If the classical limit for the clock is not taken, a fully quantum-mechanical 
  equation of motion for the system is obtained from the Exact Factorization.
  We call this equation the clock-dependent Schr\"odinger equation (CDSE) \cite{schild2018}
  because the state of the quantum system does not depend conditionally on a 
  classical time parameter, like in the TDSE, but on the state of the quantum 
  clock.
  Instead of an unobservable time parameter, only observable quantities like 
  the configuration and momentum of the clock appear in the CDSE.
  
  In this article, we investigate how the equations of quantum hydrodynamics
  change if, instead of the TDSE, the CDSE is used as starting point for its 
  derivation.
  For this purpose, in section \ref{sec:clali} we explain what is meant with 
  the classical limit and in section \ref{sec:tdqh} we provide a brief review 
  of the basis of time-dependent quantum hydrodynamics and explain how the 
  trajectory-based picture of Bohmian mechanics can be derived.
  Thereafter, in section \ref{sec:cdse} the CDSE is introduced which is used, 
  in section \ref{sec:cdqh}, to derive clock-dependent quantum hydrodynamics
  and a clock-dependent trajectory picture.
  To illustrate these trajectories, a simple model of an electron dynamics 
  with a quantum nucleus and an external time parameter is presented in 
  section \ref{sec:example}.
  Finally, in section \ref{sec:discussion} some implications and possible 
  applications of clock-dependent quantum hydrodynamics are discussed. 
  
  The notation in the following is such that similar quantities are denoted 
  with similar symbols.
  However, to avoid notational clutter the symbols may be re-defined in 
  different parts of the text.
  Care is taken to define the symbols correctly, but the preceding definitions 
  need to be considered if equations of different parts of the text are to be 
  compared.
  In particular, same symbols with different dependencies may refer to different
  (but related) quantities.
  
  
  \section{The classic limit}
  \label{sec:clali}
  
  Below, in several occasions a classical limit of quantum mechanics is 
  mentioned.
  In this section, it is briefly explained in which sense this classical limit 
  is to be understood here (see e.g.\ \cite{klein2012} for a critical discussion).
  The evolution of a single-particle quantum system described by its state 
  function $\varphi(\vec{x}|t) \in \mathbb{C}$ is given by the TDSE
  \begin{align}
    i \hbar \pt \varphi(\vec{x}|t) = \left(\frac{\pop^2}{2 m} + V(\vec{x},t) \right) \varphi(\vec{x}|t)
    \label{eq:tdse}
  \end{align}
  where $\pop = -i \hbar \px$ is the momentum operator, $m$ is the mass of 
  the particle, $t \in \mathbb{R}$ is the time parameter, $\pt$ is the 
  derivative w.r.t.\ $t$, $\vec{x} \in \mathbb{R}^3$ is the position of the 
  particle and $\px$ is the 
  gradient w.r.t.\ $\vec{x}$.
  For simplicity, we assume that there is only a scalar potential $V(\vec{x},t)$
  that represents the interaction with the environment of the particle, and 
  that no vector potential is present.
  We write 
  \begin{align}
    \varphi(\vec{x}|t) = e^{i z(\vec{x}|t) / \hbar}
  \end{align}
  and expand the action $z$ in powers of $\hbar$,
  \begin{align}
    z(\vec{x}|t) &= \sum_{j=0}^{\infty} \hbar^j z^{(j)}(\vec{x}|t).
  \end{align}
  In principle, for such an expansion a dimensionless parameter should be 
  chosen instead of $\hbar$.
  The difference between using such a parameter and using $\hbar$ seems to be 
  largely irrelevant in single-particle quantum mechanics but can be important for 
  multi-particle systems where the classical limit is only made for a part 
  of the total system.
  Then, a suitable parameter is needed which effectively becomes a pre-factor for 
  those appearances of $\hbar$ that actually relate to the part for which the 
  classical limit is made.
  Below, such a partial classical limit is discussed for a clock as part of 
  a quantum super-system, but for simplicity we limit the discussion here to 
  an expansion in terms of $\hbar$. 
  Details about how to perform the classical limit correctly for the case 
  discussed below can be found in \cite{eich2016}.
  
  To lowest order in $\hbar$, we find
  \begin{align}
    0 &= \pt s + \frac{\left( \px s \right)^2}{2m} + V 
    \label{eq:claslim_t}
  \end{align}
  with $s(\vec{x}|t) \equiv z^{(0)} \in \mathbb{R}$ being real-valued.
  This is a classical Hamilton-Jacobi equation with classical momentum $\px s$,
  hence it can be viewed as classical limit of the quantum problem.
  Interestingly, as $s$ is (chosen) real-valued, the wavefunction to lowest 
  order in $\hbar$ corresponds to a constant probability density, $|\varphi^{(0)}(\vec{x}|t)|^2 = |e^{i s}|^2 = 1$.
  The quantum properties of the system arise from a confinement (due to 
  interaction with other particles) which is represented by a variation of the 
  density.
  We find the correction to first order in $\hbar$ from 
  \begin{align}
    \pt w
      &= - \frac{\px s \px w}{m}-\frac{\px^2 s}{2m}
  \end{align}
  with $w(\vec{x}|t) = i z^{(1)} \in \mathbb{R}$.
  This correction provides such a variation of the density, 
  $|\varphi^{(1)}(\vec{x}|t)|^2 = |e^{w + i s}|^2 = e^{2 w}$.
  
  To conclude the short discussion of the classical limit, we note that if we 
  had started from a TISE
  \begin{align}
    E \varphi(\vec{x}|t) = \left(\frac{\pop^2}{2 m} + V(\vec{x},t) \right) \varphi(\vec{x}|t)
    \label{eq:tise}
  \end{align}
  with energy $E$, the classical limit would correspond to the time-independent 
  Hamilton-Jacobi equation 
  \begin{align}
    E &= \frac{(\px s)^2}{2m} + V.
    \label{eq:tihj}
  \end{align}
  This classical limit is needed below because we discuss a closed (and hence 
  time-independent) super-system to define time internally as correlation of 
  a part of the super-system to another part, which serves as the clock.
  The trajectories of a classical clock are then given as solutions of \eqref{eq:tihj}.

  
  \section{Time-dependent quantum hydrodynamics}
  \label{sec:tdqh}
  
  The hydrodynamic formulation of quantum mechanics \cite{madelung1926,madelung1927} 
  for a single-particle system is obtained from the TDSE \eqref{eq:tdse} by writing the 
  complex number $\varphi$ in its polar form,
  \begin{align}
    \varphi(\vec{x}|t) 
      &= \sqrt{\rho(\vec{x}|t)} e^{i s(\vec{x}|t)}        \label{eq:varphi_polar}  \\
      &= e^{w(\vec{x}|t) + i s(\vec{x}|t)} \label{eq:varphi_polar2}
  \end{align}
  with probability density $\rho(\vec{x}|t) = |\varphi(\vec{x}|t)|^2 \in \mathbb{R}$ 
  and with real-valued dimensionless action $s \in \mathbb{R}$.
  In \eqref{eq:varphi_polar2}, also a second real-valued dimensionless 
  action $w \in \mathbb{R}$ is introduced which may be used instead of the 
  probability density $\rho$.
  By inserting \eqref{eq:varphi_polar} into the TDSE \eqref{eq:tdse} and by 
  separating the result into real and imaginary parts, two equations are 
  obtained.
  The first of these equations is a Hamilton-Jacobi equation,
  \begin{align}
    0 = \hbar \pt s(\vec{x}|t) + H(\vec{p},\vec{x}|t),
    \label{eq:tdhje}
  \end{align}
  where
  \begin{align}
    H(\vec{p},\vec{x}|t) &= \frac{\vec{p}(\vec{x}|t)^2}{2 m} + V(\vec{x}|t) + u(\vec{x}|t)
  \end{align}
  is a Hamiltonian function,
  \begin{align}
    \vec{p}(\vec{x}|t) = \hbar \px s(\vec{x}|t)
    \label{eq:p}
  \end{align}
  is a real-valued momentum field, and $u(\vec{x}|t)$ is the so-called quantum 
  potential
  \begin{align}
    u(\vec{x}|t) = -\frac{1}{2 m} \left( \popc \cdot \vec{\pi}(\vec{x}|t) + \vec{\pi}(\vec{x}|t)^2 \right)
    \label{eq:qpot}
  \end{align}
  that vanishes if a suitable classical limit is taken for $\varphi(\vec{x}|t)$.
  In the expression for the quantum potential \eqref{eq:qpot}
  we used the operator 
  \begin{align}
   \popc &= i \pop = \hbar \px
   \label{eq:popc}
  \end{align}
  and the quantity
  \begin{align}
    \vec{\pi}(\vec{x}|t) = \frac{\hbar \px \rho}{2 \rho} = \hbar \px w(\vec{x}|t),
  \end{align}
  which is another real-valued momentum field, $\vec{\pi}\in \mathbb{R}$ (also
  used e.g.\ in \cite{garashchuk2014,gao2017}).
  It represents the relative variation $\frac{\px \rho}{\rho}$ of 
  the density with position $\vec{x}$. 
  The second of the equations obtained from \eqref{eq:varphi_polar} and 
  \eqref{eq:tdse} is the continuity equation 
  \begin{align}
    0 = \pt \rho(\vec{x}|t) + \px \cdot \vec{j}(\vec{x}|t)
    \label{eq:tdce}
  \end{align}
  with probability flux (current) density
  \begin{align}
    \vec{j}(\vec{x}|t) = \rho(\vec{x}|t) \frac{\vec{p}(\vec{x}|t)}{m}.
    \label{eq:fd}
  \end{align}
  The continuity equation \eqref{eq:tdce} represents the conservation of the 
  probability density $\rho(\vec{x}|t)$.
  It essentially states that in any volume $\Omega$, the change of density 
  $\pt \rho(\vec{x}|t)$ is given by the flow of density through the boundary 
  of $\Omega$, and that flow is determined by the probability 
  flux density $\vec{j}(\vec{x}|t)$.\cite{aris1989}
  For comparison with the development presented below, it is also useful to 
  express \eqref{eq:tdce} with the momentum field $\vec{\pi}$, i.e., the 
  continuity equation is equivalent to
  \begin{align}
    0 &= \hbar^2 \pt w + \frac{1}{m} \left( \vec{\pi} \cdot \vec{p} + \frac{\popc \cdot \vec{p}}{2} \right)
       = \hbar^2 \pr w + \frac{\px \cdot \vec{j}}{2 \rho}.
    \label{eq:tdce1}
  \end{align}
  While version \eqref{eq:tdce} of the continuity equation contains quantities 
  that are non-zero only in regions where the particle actually can be found 
  (in the sense of its probability distribution), version \eqref{eq:tdce1} relates 
  quantities that can be sizable everywhere.
  
  To obtain the particle-based interpretation of quantum mechanics known as 
  Bohmian mechanics, it is necessary to solve the Hamilton-Jacobi equation
  \eqref{eq:tdhje} with the method of characteristics.\cite{agostini2018}
  By using this method, \eqref{eq:tdhje} is interpreted as differential 
  equation for the phase $s(\vec{x}|t)$ alone, i.e., it is assumed that the quantum 
  potential $u(\vec{x}|t)$ (or the density $\rho(\vec{x}|t)$, or $w(\vec{x}|t)$) 
  is known.
  Then, \eqref{eq:tdhje} is solved along parametrized curves.
  It turns out that the parameter can be identified with $t$ and that the 
  curves can be obtained by solving 
  \begin{align}
    \pt \vec{x}_{\rm t}(t) &= \frac{\vec{p}_{\rm t}(t)}{m} 
      \label{eq:traj_x} \\
    \pt \vec{p}_{\rm t}(t) &= -\px H(\vec{p}_{\rm t}(t),\vec{x}_{\rm t}(t)|t) 
      \label{eq:traj_p}\\
    \pt S_{\rm t}(t)       &= \frac{\left( \vec{p}_{\rm t}(t) \right)^2}{m} - H(\vec{p}_{\rm t}(t),\vec{x}_{\rm t}(t)|t).
      \label{eq:traj_S}
  \end{align}
  Here, $\vec{x}_{\rm t}$, $\vec{p}_{\rm t}$, and $S_{\rm t}$ are the values
  of $\vec{x}$, $\vec{p}$, and $S$ along the trajectory, respectively.
  
  From \eqref{eq:traj_x} it is possible to interpret the curves $\vec{x}_{\rm t}(t)$ 
  as trajectories of the particle which are guided by the wavefunction 
  $\varphi(\vec{x}|t)$ via the momentum field $\vec{p}(\vec{x}|t)$ or, 
  if the momentum field is computed from \eqref{eq:traj_p}, via the quantum 
  potential $u(\vec{x}|t)$ appearing in the Hamiltonian function $H$.
  If several trajectories are considered with their initial locations 
  $\vec{x}_{\rm t}(t_0)$ randomly sampled according to $\rho(\vec{x}|t_0)$ at 
  some time $t_0$, or if the trajectories are equally spaced but carry a weight 
  according to this distribution, the continuity equation \eqref{eq:tdce} 
  ensures that the trajectories yield the distribution $\rho(\vec{x}|t)$ for 
  any time $t$.
  Hence, the basic equations of quantum hydrodynamics may be interpreted as giving
  rise to a particle picture of quantum mechanics where the particles have a 
  definite trajectory in space, but the trajectories are guided by the phase of 
  the wavefunction (or the quantum potential) and distributed according to its 
  squared magnitude.
  
  It has to be noted that a few unusual conventions were chosen in the 
  preceding discussion of time-dependent quantum hydrodynamics.
  First, it is custom to divide $w$ and $s$ by $\hbar$ in the ansatz
  \eqref{eq:varphi_polar}, so as to give them units of action.
  We made this choice above when discussion the classical limit.
  It is convenient for single-particle quantum mechanics but less convenient for 
  the treatment presented below where vector potentials appear, hence it is not 
  done here.
  
  Second, momentum fields are considered, which is typically done for the 
  Hamilton-Jacobi equation but which is in contrast to much literature on 
  quantum hydrodynamics and Bohmian mechanics, where a formulation of the 
  equations in terms of velocity fields tends to be preferred.
  The preference might originate from the fact that the velocity field 
  $\frac{\vec{p}(\vec{x}|t)}{m}$ occurs in the definition of the flux density 
  $\vec{j}(\vec{x}|t)$ \eqref{eq:fd} and in the guiding equation \eqref{eq:traj_x}
  of Bohmian mechanics, and it also prevents confusion with the momentum 
  operator $\pop$.
  That momentum fields are considered here is because of this authors subjective 
  choice, but a reformulation in terms of velocity fields is straightforward.
  
  Third, the quantum potential is usually given as
  \begin{align}
    u(\vec{x}|t) = -\frac{\hbar^2}{2 m} \frac{\px^2 |\varphi|}{|\varphi|}
  \end{align}
  and the momentum field $\vec{\pi}(\vec{x}|t)$ is not introduced.
  However, if time is replaced by a quantum clock, it is more convenient to 
  work with the two momentum fields $\vec{p}$ and $\vec{\pi}$ than with other 
  quantities like the density $\rho$ and flux density $\vec{j}$ in the sense 
  that the equations become more transparent, even though the densities have 
  the virtue of being of relevant magnitude only in regions where the particle 
  can actually be found.
  In the classical limit corresponding to \eqref{eq:claslim_t}, the momentum 
  field $\vec{p}$ becomes the classical momentum while the momentum field 
  $\vec{\pi}$ (and hence the quantum potential $U$) vanish.
  We will thus call $\vec{p}$ the classical momentum field and $\vec{\pi}$ the 
  quantum momentum field in the following.
  Also, using $\vec{\pi}(\vec{x}|t)$ makes the quantum potential look 
  (partially) like an additional kinetic energy term.
  As explained in section \ref{sec:clali}, the variation of the density 
  (which is what $\vec{\pi}$ represents) is a sign of ``quantum behavior'' of 
  the system, but it also reflects an effective interaction with other particles
  that are not treated explicitly but only implicitly via a scalar potential 
  $V$ or a vector potential.
  Hence, the quantum potential may loosely be interpreted as effective reaction 
  of the system on its confinement.
  
  
  
  \section{The clock-dependent Schr\"odinger equation}
  \label{sec:cdse}
  
  As stated above, the TDSE is a semi-classical description of a quantum system 
  because its time parameter originates from an implicit comparison of the state 
  of the quantum system to the state of a classical clock.
  A quantum mechanical generalization of the TDSE, the CDSE, can be obtained as 
  follows:\cite{schild2018}
  The state $\psi(\vec{R},\vec{r})$ of a super-system composed of two parts
  is determined by the TISE 
  \begin{align}
    \left( \frac{\Pop^2}{2M} + \frac{\pop^2}{2m} + V(\vec{R},\vec{r}) \right) \psi(\vec{R},\vec{r}) = E \psi(\vec{R},\vec{r}),
    \label{eq:tise_psi}
  \end{align}
  where $\Pop = -i \hbar \pR$ and $\pop = -i \hbar \pr$ are two momentum 
  operators and $V(\vec{R},\vec{r})$ is a scalar potential.
  This equation is written for two particles with coordinates $\vec{R}$ and 
  $\vec{r}$ and masses $M$ and $m$, respectively, but it can be generalized to 
  any number of particles by replacing the two kinetic energy operators that 
  occur in \eqref{eq:tise_psi} with sums of such operators.
  Also, the momentum operators $\Pop$ and $\pop$ may include vector potentials
  without changing the results discussed below in any relevant way.
  Next, the joint probability density $|\psi(\vec{R},\vec{r})|^2$ is written 
  as product
  \begin{align}
    |\psi(\vec{R},\vec{r})|^2 = |\chi(\vec{R})|^2 |\phi(\vec{r}|\vec{R})|^2,
  \end{align}
  where 
  \begin{align}
    |\chi(\vec{R})|^2 := \Braket{\psi(\vec{R},\vec{r})|\psi(\vec{R},\vec{r})}
    \label{eq:chi}
  \end{align}
  is the marginal probability density of finding the particle of mass $M$ at 
  $\vec{R}$ independent of where the particle of mass $m$ is.
  The symbol $\Braket{\cdot}$ indicates integration over the coordinates 
  $\vec{r}$ and $\Braket{a|b} = \Braket{\bar{a} b}$ is the complex-valued 
  scalar product, where $\bar{a}$ is the complex-conjugate of $a$.
  The function $\chi(\vec{R})$ is called marginal amplitude or marginal 
  wavefunction and its phase can be chosen arbitrarily (which leads to a gauge 
  freedom in the theory, as discussed below).
  The probability density $|\phi(\vec{r}|\vec{R})|^2$ is the conditional 
  probability density for finding the particle with mass $m$ at $\vec{r}$ 
  given the particle with mass $M$ is at $\vec{R}$.
  The conditional amplitude or conditional wavefunction is defined as 
  \begin{align}
    \phi(\vec{r}|\vec{R}) := \frac{\psi(\vec{R},\vec{r})}{\chi(\vec{R})}
  \end{align}
  and, if $\psi(\vec{R},\vec{r})$ is normalized according to
  \begin{align}
    \int \Braket{\psi(\vec{R},\vec{r}) | \psi(\vec{R},\vec{r})} d\vec{R} \stackrel{!}{=} 1,
  \end{align}
  it needs to obey the partial normalization condition
  \begin{align}
    \Braket{\phi(\vec{r}|\vec{R}) | \phi(\vec{r}|\vec{R})} \stackrel{!}{=} 1
  \end{align}
  for all $\vec{R}$, as is required for a conditional probability.
  Then, the marginal amplitude is normalized as
  \begin{align}
    \int |\chi(\vec{R})|^2 d\vec{R} = 1.
  \end{align}
  
  The idea to write the wavefunction as a product of a marginal and a 
  conditional wavefunction was brought up some time ago \cite{hunter1975} and has
  recently been developed further under the name Exact Factorization.\cite{abedi2010,abedi2012}
  The equations of motion for $\chi(\vec{R})$ and $\phi(\vec{r}|\vec{R})$ that 
  follow are
  \begin{align}
    E       \chi &= \left( \frac{(\Pop + \vec{A}(\vec{R}))^2}{2M} + \epsilon(\vec{R}) \right) \chi \\
    \hat{C} \phi &= \left( \frac{\pop^2}{2m} + V(\vec{R},\vec{r}) + \hat{U} - \epsilon(\vec{R}) \right) \phi \label{eq:phi}
  \end{align}
  with scalar and vector potentials
  \begin{align}
    \epsilon(\vec{R}) &:= \Braket{\phi|\frac{\pop^2}{2m} + V(\vec{R},\vec{r}) + \hat{U} - \hat{C}|\phi} \\
    \vec{A}           &:= \Braket{\phi|\Pop \phi},
  \end{align}
  with the kinetic operator
  \begin{align}
    \hat{U} &= \frac{(\Pop-\vec{A})^2}{2M}
  \end{align}
  and with the coupling operator
  \begin{align}
    \hat{C} &= -\frac{1}{M} \frac{(\Pop+\vec{A}) \chi(\vec{R})}{\chi(\vec{R})} \cdot \left( \Pop - \vec{A} \right)
  \end{align}
  that depends explicitly on the wavefunction $\chi(\vec{R})$.
  The choice of the phase of $\chi(\vec{R})$ is a gauge freedom, i.e., the 
  transformation 
  \begin{align}
    \chi'(\vec{R})         &= \chi(\vec{R}) e^{-i \theta(\vec{R})}        \\
    \phi'(\vec{r}|\vec{R}) &= \phi(\vec{r}|\vec{R}) e^{i \theta(\vec{R})} \\
    \vec{A}'(\vec{R})      &= \vec{A}(\vec{R}) + \pR \theta(\vec{R})
  \end{align}
  leaves the total wavefunction $\psi(\vec{R},\vec{r})$ as well as all 
  equations of motion in the theory invariant.
  
  The marginal wavefunction $\chi(\vec{R})$ is interpreted as the wavefunction 
  of the clock or environment, whereas the conditional wavefunction 
  $\phi(\vec{r}|\vec{R})$ is the wavefunction of the quantum system of 
  interest.
  It has been shown that if the clock behaves classical, the equation of motion 
  for $\chi(\vec{R})$ becomes the analogue of the time-independent 
  Hamilton-Jacobi equation \eqref{eq:tihj}.\cite{briggs2000,briggs2015,eich2016,schild2018}
  Solving this equation for its characteristics yields classical trajectories.
  The corresponding classical configuration and momentum of the clock along these 
  trajectories can be parametrized with a variable $t$ which may be interpreted 
  as time parameter.
  In the equation of motion for $\phi(\vec{r}|\vec{R})$, the contribution of the 
  operator $\hat{U}$ vanishes, the clock-dependent operator becomes
  \begin{align}
    \hat{C} \rightarrow i \hbar \vec{V}_{\rm cl} \nabla_{\vec{R}_{\rm cl}} \rightarrow i \hbar \pt
  \end{align}
  with classical configuration $\vec{R}_{\rm cl}(t)$ and velocity 
  $\vec{V}_{\rm cl}(t) = \pt \vec{R}_{\rm cl}(t)$ of the clock,
  and \eqref{eq:phi} turns into a normal TDSE for the conditional subsystem.
  This is what is meant with the ``classical limit'' in figure \ref{fig:scheme}.
  Thus, the determining equation for the conditional system \eqref{eq:phi} can 
  be considered the quantum-mechanical analogue of the TDSE -- the CDSE -- 
  where the clock that is used to define the time parameter is treated fully 
  quantum-mechanically.
  
  
  \section{Clock-dependent quantum hydrodynamics}
  \label{sec:cdqh}
  
  The basic equations of time-dependent quantum hydrodynamics, i.e., the 
  continuity equation \eqref{eq:tdce} and the Hamilton-Jacobi equation 
  \eqref{eq:tdhje}, are derived from the TDSE. 
  Similarly, the corresponding clock-dependent equations can be found from the 
  CDSE. 
  The derivation proceeds along the same lines as for the time-dependent case,
  i.e., the conditional wavefunction $\phi(\vec{r}|\vec{R})$ is written in its 
  polar form and inserted into the CDSE, the equation is separated into its 
  real and imaginary parts, and the resulting equations are expressed in terms 
  of the probability density $|\phi(\vec{r}|\vec{R})|^2$, momentum densities,
  and probability flux densities.
  Before giving the results of this procedure, a few quantities need to be 
  defined.
  
  For a generic function $\varphi(\vec{R},\vec{r}) = e^{w(\vec{R},\vec{r}) + i s(\vec{R},\vec{r})} \in \mathbb{C}$ with 
  $s, w \in \mathbb{R}$, two momentum fields w.r.t.\ $\vec{R}$ are defined.
  The first is the classical momentum field
  \begin{align}
    \vec{P}[\varphi,\vec{A}] 
      :=\hbar \frac{\operatorname{Im} (\bar{\varphi} \pR \varphi)}{|\varphi|^2} + \vec{A}
       = \hbar \pR s + \vec{A}
  \end{align}
  and the second is the quantum momentum field
  \begin{align}
    \vec{\Pi}[\varphi] 
      := \hbar \frac{\operatorname{Re} (\bar{\varphi} \pR \varphi)}{|\varphi|^2}
      = \hbar \pR w.
  \end{align}
  Both fields are real-valued, $\{\vec{P}, \vec{\Pi}\} \in \mathbb{R}$.
  The flux density corresponding to $\vec{P}$ is
  \begin{align}
    \vec{J}[\varphi,\vec{A}] &= \frac{1}{M} |\varphi|^2 \vec{P}[\varphi,\vec{A}].
  \end{align}
  Similar fields are defined w.r.t.\ $\vec{r}$, i.e., the classical momentum field
  \begin{align}
    \vec{p}[\varphi] 
      := \hbar \frac{\operatorname{Im} (\bar{\varphi} \pr \varphi)}{|\varphi|^2}
       = \hbar \pr s
  \end{align}
  and the quantum momentum field
  \begin{align}
    \vec{\pi}[\varphi] 
      := \hbar \frac{\operatorname{Re} (\bar{\varphi} \pr \varphi)}{|\varphi|^2}
      = \hbar \pr w,
  \end{align}
  with $\{ \vec{p}, \vec{\pi} \} \in \mathbb{R}$.
  The corresponding flux density is 
  \begin{align}
    \vec{j}[\varphi] &= \frac{1}{m} |\varphi|^2 \vec{p}[\varphi].
  \end{align}
  Finally, we define the probability density for the system depending on 
  $\vec{r}$ and conditionally on $\vec{R}$ as
  \begin{align}
    \rho(\vec{r}|\vec{R}) = |\phi(\vec{r}|\vec{R})|^2.
  \end{align}
  
  With all these definitions, the clock-dependent Hamilton-Jacobi equation (CDHJE)
  can be written in the compact form
  \begin{align}
    0 =& \frac{1}{M} 
        \left( \vec{P}[\chi,\vec{A}] \cdot \vec{P}[\phi,-\vec{A}] 
             - \vec{\Pi}[\chi] \cdot \vec{\Pi}[\phi] \right) 
    - \epsilon(\vec{R})\nonumber \\
      & + H_{\rm S }(\vec{r}|\vec{R}) 
        + H_{\rm SC}(\vec{r}|\vec{R}) 
    \label{eq:cdhje}
  \end{align}
  where we defined the Hamiltonian function for the system w.r.t.\ the system coordinates,
  \begin{align}
    H_{\rm S }(\vec{r}|\vec{R}) 
      &:= \frac{\left(\vec{p}[\phi]\right)^2}{2 m} + u(\vec{R},\vec{r}) + V(\vec{R},\vec{r})
  \end{align}
  and the Hamiltonian function for the system, but w.r.t.\ the clock coordinates,
  \begin{align}
    H_{\rm SC}(\vec{r}|\vec{R})
      &:= \frac{\left(\vec{P}[\phi,-\vec{A}]\right)^2}{2 M} + U(\vec{R},\vec{r}).
  \end{align}
  These Hamiltonian functions contain the quantum potentials w.r.t.\ the 
  coordinates of the system,
  \begin{align}
    u(\vec{R},\vec{r}) = -\frac{1}{2m} \left( \popc \cdot \vec{\pi}[\phi] + \left(\vec{\pi}[\phi]\right)^2 \right),
  \end{align}
  and w.r.t.\ the coordinates of the clock,
  \begin{align}
    U(\vec{R},\vec{r}) = -\frac{1}{2M} \left( \Popc \cdot \vec{\Pi}[\phi] + \left(\vec{\Pi}[\phi]\right)^2 \right).
  \end{align}
  As in \eqref{eq:popc} of the treatment of time-dependent quantum hydrodynamics, 
  the momentum operators appearing in these quantum potentials are the 
  real-valued analogue of the usual momentum operators, i.e., 
  \begin{align}
    \popc &= i \pop = \hbar \pr \\
    \Popc &= i \Pop = \hbar \pR
  \end{align}
  
  Although \eqref{eq:cdhje} looks rather different than its time-dependent 
  version \eqref{eq:tdhje}, there is some similarity.
  In particular, we have the correspondence
  \begin{align}
    \frac{1}{M} \vec{P}[\chi,\vec{A}] \cdot \vec{P}[\phi,-\vec{A}] \leftrightarrow \hbar \pt S(\vec{x}|t) \label{eq:cor1} \\
    H_{\rm S }(\vec{p}[\phi],\vec{r}|\vec{R})          \leftrightarrow H(\vec{p},\vec{x}|t).
  \end{align}
  Instead of the time-derivative we have the scalar product 
  $\vec{P}[\chi,\vec{A}] \cdot \vec{P}[\phi,-\vec{A}]$.
  It contains the derivative of the phase of $\phi$ w.r.t.\ the clock 
  configuration.
  For $\vec{A} = \vec{0}$, we have that correspondence \eqref{eq:cor1} is 
  just 
  \begin{align}
    \frac{1}{M} \vec{P}[\chi] \cdot \pR \leftrightarrow \pt.
  \end{align}
  Some additional terms also appear in the fully quantum-mechanical equation 
  \eqref{eq:cdhje} that have no equivalent in the time-dependent Hamilton-Jacobi
  equation \eqref{eq:tdhje}:
  Those are the term $\vec{\Pi}[\chi] \cdot \vec{\Pi}[\phi]$, which connects the 
  quantum momentum fields of system and clock w.r.t.\ the configuration of the 
  clock, and the Hamiltonian function $H_{\rm SC}$, which contains a kinetic 
  energy term and a quantum potential of the system w.r.t.\ the configuration 
  of the clock.
  
  The clock-dependent continuity equation (CDCE) is 
  \begin{align}
    0 =& \frac{1}{M} \vec{P}[\chi,\vec{A}] \cdot \pR \rho
       + \pr \cdot \vec{j}_{\rm C}[\phi] \nonumber \\
      &+ \pR \cdot \vec{J}_{\rm C}[\phi,-\vec{A}]
       + \frac{2 i}{\hbar} \vec{\Pi}[\chi] \cdot \vec{J}_{\rm C}[\phi,-\vec{A}].
       \label{eq:cdce}
  \end{align}
  It was already introduced and discussed in \cite{schild2018}, and it is 
  similar to its time-dependent counterpart \eqref{eq:tdce} in the sense of 
  the correspondences
  \begin{align}
    \frac{1}{M} \vec{P}[\chi,\vec{A}] \cdot \pR \rho(\vec{r}|\vec{R}) & \leftrightarrow \pt \rho(\vec{x}|t) \\
    \pr \cdot \vec{j}_{\rm C}[\phi](\vec{r}|\vec{R}) & \leftrightarrow \px \cdot \vec{j}(\vec{x}|t).
  \end{align}
  The terms including the flux density $\vec{J}_{\rm C}[\phi,-A]$ w.r.t.\ the 
  configuration of the clock appear only in the clock-dependent treatment.
  
  The continuity equation \eqref{eq:cdce} was written such that it resembles 
  its time-dependent counterpart \eqref{eq:tdce} as closely as possible.
  However, a more natural way of writing \eqref{eq:cdce} is
  \begin{align}
    0 =& \frac{1}{M} \left( \vec{P}[\chi,\vec{A}] \cdot \vec{\Pi}[\phi]
                          + \vec{\Pi}[\chi] \cdot \vec{P}[\phi,-\vec{A}]
                \right) \nonumber \\
      &+ \frac{1}{m} \left( \vec{\pi}[\phi] \cdot \vec{p}[\phi] + \frac{\popc \cdot \vec{p}[\phi]         }{2} \right) \nonumber \\
      &+ \frac{1}{M} \left( \vec{\Pi}[\phi] \cdot \vec{P}[\phi,-\vec{A}] + \frac{\Popc \cdot \vec{P}[\phi,-\vec{A}]}{2} \right),
      \label{eq:cdce1}
  \end{align}
  which is entirely in terms of the momentum fields (compare to \eqref{eq:tdce1}
  for the time-dependent case).
  Writing the continuity equation as \eqref{eq:cdce1} allows to compare it 
  directly with the clock-dependent Hamilton-Jacobi equation \eqref{eq:cdhje}.
  It is apparent that in the latter, only products of two classical momentum 
  fields ($\vec{p}[\dots]$ or $\vec{P}[\dots]$) or two quantum momentum fields 
  ($\vec{\pi}[\dots]$ or $\vec{\Pi}[\dots]$) appear, as well as the second 
  derivatives of the quantum momentum fields.
  In contrast, in the clock-dependent continuity equation only mixed products 
  of one classical and one quantum momentum field are found, as well as the 
  second derivatives of the classical momentum fields.
  
  As the clock-dependent Hamilton Jacobi equation \eqref{eq:cdhje} can
  be interpreted as differential equation for the phase $\arg(\phi)$ alone, it 
  can be solved in terms of trajectories with the method of characteristics.
  Given the momentum fields, the trajectories are obtained by solving 
  \begin{align}
    \partial_{\tau} \vec{r}_{\rm t}(\tau)    
      =& \frac{\vec{p}_{\rm t}[\phi]}{m}   \label{eq:pt1}\\
    \partial_{\tau} \vec{R}_{\rm t}(\tau)    
      =& \frac{\vec{P}_{\rm t}[\chi,A] + \vec{P}_{\rm t}[\phi,-A]}{M} \equiv \frac{\vec{P}_{\rm t}[\psi]}{M} \label{eq:pt2}
  \end{align}
  where $\tau$ is the parameter and the subscript ``${\rm t}$'' means that the 
  corresponding quantity is evaluated along the trajectory.
  The parameter $\tau$ is an arbitrary parameter, but it very much reminds of 
  the time parameter.
  However, $\vec{r}_{\rm t}$ depends conditionally on the actual trajectory 
  $\vec{R}_{\rm t}$ of the clock via the conditional dependence of the 
  momentum field $\vec{p}[\phi]$.
  The position of the clock, $\vec{R}_{\rm t}$, is determined by the momentum 
  field of both the clock and the system w.r.t.\ the clock coordinates, which 
  corresponds to the (gauge-invariant) momentum field of the state $\psi$ of 
  the super-system w.r.t.\ the clock coordinates.

  
  \section{Application to a model of an electron dynamics with quantum nuclei}
  \label{sec:example}
  
  To illustrate clock-dependent quantum hydrodynamics, the model of \cite{eich2016}
  for a proton-coupled electron transfer is used which has also previously been 
  studied in connection with the clock-dependent continuity equation.\cite{schild2018}
  It is a model for a dynamics of an electronic and a nuclear degree of freedom, 
  where the dynamics itself is generated from a TDSE that refers to an external 
  time parameter $t$.
  Hence, the model is a model for time- and clock-dependent quantum hydrodynamics 
  with the quantum system being the electronic degree of freedom, the quantum 
  clock being the nuclear degree of freedom, and with time $t$ defined by an
  unspecified but essentially classical clock, e.g.\ a laser field that 
  initiates the dynamics and that might be used to also probe the dynamics.
  The model was chosen for two reasons:
  First, it is not straightforward to find a model for clock-dependent quantum 
  hydrodynamics alone which is non-trivial.
  This hints at that there is something missing in the theory, which may be 
  called ``the reason why there is dynamics''.
  Second, trajectory-based time- and clock-dependent quantum hydrodynamics can
  be a useful method for calculating a complicated quantum dynamics where quantum 
  effects need to be treated but a full solution in terms of wavefunctions is not 
  feasible.
  Both points are further discussed in section \ref{sec:discussion}.
  
  \begin{figure}[htbp]
    \centering
    \includegraphics[width=0.99\textwidth]{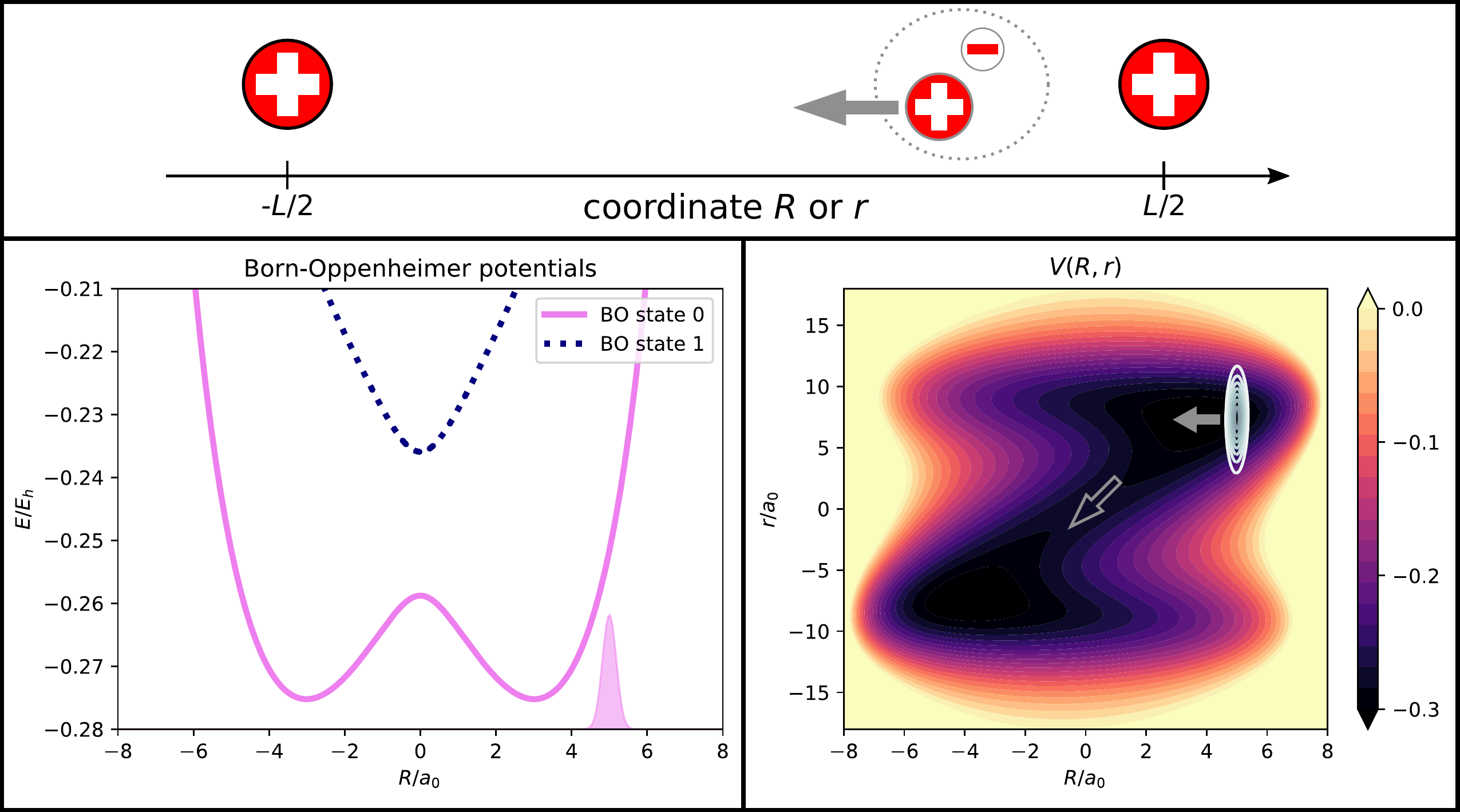}
    \caption{Top: A heavy particle with positive charge is positioned along 
             coordinate $R$ and a light particle with negative charge is 
             positioned along coordinate $r$, while two positive charges are 
             clamped at $R = \pm L$, $r = \pm L$. The parameters of the model are 
             chosen such that the positive charge gains momentum towards 
             smaller $R$, dragging the negative charge along smaller $r$.
             Bottom left: The lowest two Born-Oppenheimer potentials for the 
             heavy positive charge. Also shown is the initial marginal density 
             of the positive charge. Bottom right: Model potential $V(R,r)$ and 
             initial wavefunction for the dynamics. Two arrows indicate the 
             motion during the dynamics.
             }
    \label{fig:potentials}
  \end{figure}

  A sketch of the model is depicted at the top of figure \ref{fig:potentials}.
  A heavy positively charged particle can move along coordinate $R$ (the nucleus),
  and a light negatively charged particle can move along coordinate $r$ (the 
  electron).
  Two clamped (infinitely heavy) positive charges are located at 
  $(R,r) = (\pm L/2, \pm L/2)$.
  The model Hamiltonian is
  \begin{align}
    H = -\frac{\mu}{2} \partial_R^2 + \hat{H}_{\rm S} 
  \end{align}
  with $\mu = m/M$ being the mass ratio between a light negatively charged 
  particle (the electron) moving along dimension $r$ and a heavy positively 
  charged particle (the nucleus) moving along dimension $R$, where
  \begin{align}
    \hat{H}_{\rm S} = -\frac{\pr^2}{2} + V(R,r)
    \label{eq:hs}
  \end{align}
  contains the kinetic energy of the light particle and the scalar interaction
  potential
  \begin{align}
    V(R,r) =  \frac{1}{|R - \frac{L}{2}|}
                              + \frac{1}{|R + \frac{L}{2}|}
                              - \frac{\operatorname{erf} \left( \frac{|r-R|}{R_{\rm c}} \right)}{|R - r|}
                              - \frac{\operatorname{erf} \left( \frac{|r-\frac{L}{2}|}{R_{\rm r}} \right)}{|r - \frac{L}{2}|}
                              - \frac{\operatorname{erf} \left( \frac{|r+\frac{L}{2}|}{R_{\rm l}} \right)}{|r + \frac{L}{2}|}.
                              \label{eq:vrr}
  \end{align}
  We use the parameters \unit[$L= 19$]{$a_0$}, \unit[$R_{\rm r} = R_{\rm l} = 3.5$]{$a_0$},
  \unit[$R_{\rm c} = 4.0$]{$a_0$}, and a mass ratio $\mu^{-1} = 900$.
  For these model parameters, the Born-Oppenheimer potential energy surfaces 
  $\epsilon_n^{\rm BO}(R)$ for the nucleus, given by
  \begin{align}
    \hat{H}_{\rm S} \phi_n^{\rm BO}(r|R) = \epsilon_n^{\rm BO}(R) \phi_n^{\rm BO}(r|R),
    \label{eq:hsboa}
  \end{align}
  are energetically well-separated, as shown in the bottom-left panel of figure 
  \ref{fig:potentials}.
  The dynamics is adiabatic in the sense that 
  \begin{align}
    |\phi(r|R,t)|^2 \approx |\phi_0^{\rm BO}(r|R)|^2
    \label{eq:adia}
  \end{align}
  and the dynamics of the electron is parametrized by the configuration of the 
  nucleus (the quantum clock) but only indirectly by the external time $t$.
  
  The initial state for the dynamics is chosen to be a product of a Gaussian 
  and the ground-state Born-Oppenheimer electronic wavefunction,
  \begin{align}
    \psi_0(R,r) = \chi_0(R) \phi_0^{\rm BO}(r|R),
  \end{align}
  where 
  \begin{align}
    \chi_0(R) \propto e^{-\frac{(R-R_0)^2}{4 \sigma^2}}
  \end{align}
  with \unit[$R_0 = 5$]{$a_0$} and \unit[$\sigma = 0.183$]{$a_0$}.
  The density $|\chi_0(R)|^2$ is also shown in the bottom-left panel of figure 
  \ref{fig:potentials}.
  From the figure, it is clear that this wavepacket 
  will initially move towards smaller $R$ and has enough energy to overcome the 
  barrier at $R=0$.
  On the bottom-right panel of figure \ref{fig:potentials}, the density 
  $|\psi_0(R,r)|^2$ for the initial state is shown together with the potential 
  $V(R,r)$ of \eqref{eq:vrr}.
  The dynamics is indicated by two arrows.
  For the chosen parameters it is essentially adiabatic in the sense of the 
  Born-Oppenheimer approximation, i.e., the nucleus moves from its initial 
  center at \unit[$R = 5$]{$a_0$} towards smaller $R$ and the electron 
  follows from its initial center at ca.\ \unit[$r = 8$]{$a_0$} towards 
  smaller $r$, resulting in the motion indicated by the arrows.
  
  \begin{figure}[htbp]
    \centering
    \includegraphics[width=0.99\textwidth]{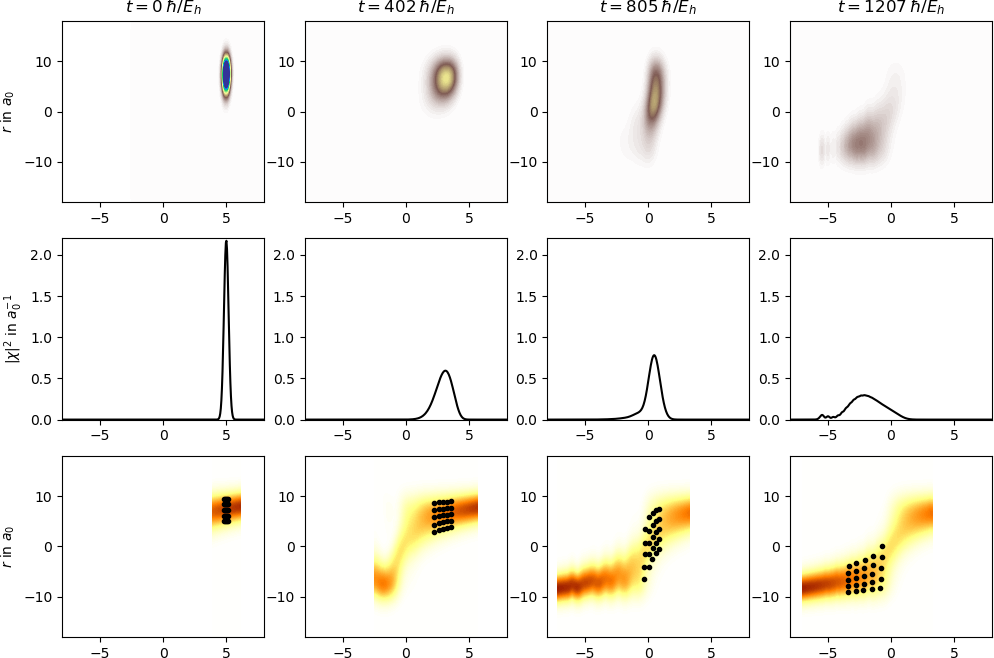}
    \caption{Density of the joint wavefunction $\psi(R,r|t)$ (top), of the 
             marginal wavefunction $\chi(R|t)$ (middle), and of the conditional 
             wavefunction $\phi(r|R,t)$ together with some of its corresponding 
             trajectories (bottom) for four different values of the external 
             time $t$.}
    \label{fig:trajectories}
  \end{figure}
  
  Figure \ref{fig:trajectories} shows the dynamics in terms of the joint 
  probability density $|\psi(R,r|t)|^2$ (top), the marginal probability density 
  of the nucleus $|\chi(R|t)|^2$ (center), and the conditional probability density 
  of the electron $|\phi(r|R,t)|^2$ (bottom), for four different values of the 
  external time $t$.
  The joint probability density follows the motion indicated by the arrows in 
  figure \ref{fig:potentials}, i.e., its maximum moves first towards smaller $R$,
  indicating the nuclear motion, but then also towards smaller $r$, indicating 
  the electron following the nucleus.
  The motion towards smaller $R$ is clearly visible in the marginal nuclear
  density $|\chi(R|t)|^2$.
  
  The probability density for our actual system of interest, the electron, is 
  the conditional density $|\phi(r|R,t)|^2$ shown as contour plot in the bottom 
  panels of figure \ref{fig:trajectories}.
  It is normalized for each value of $R$ (and $t$), hence it typically is 
  somewhere localized along $r$ for every $R$.
  In the figure, $|\phi(r|R,t)|^2$ is only shown in a region of $R$ where the 
  probability density of finding the nucleus is 
  \unit[$|\chi(R|t)|^2 > 10^{-8}$]{a$_0^{-1}$}, i.e., in a region where the 
  clock can actually be found.
  Notwithstanding, there is no principle problem in obtaining $|\phi(r|R,t)|^2$
  for all $R$, but there is a practical problem due to numerical inaccuracies
  whereever the joint probability density $|\psi(R,r|t)|^2$ becomes very small 
  in magnitude.
  
  Clock-dependent trajectories were computed for the wavefunction $\phi(r|R,t)$ 
  using \eqref{eq:pt1} and \eqref{eq:pt2}, where $\tau$ is identified with
  the external time $t$.
  The initial positions of the nucleus $R_{\rm t}(t=0)$ and of the electron
  $r_{\rm t}(t=0,R_{\rm t}(t=0))$ are chosen to be in a region where the 
  density of the joint system $|\psi(R,r|t=0)|^2$ is significant.
  Then, \eqref{eq:pt2} is solved to propagate the position of the nucleus
  $R_{\rm t}(t)$ along the trajectory.
  Thereafter, \eqref{eq:pt1} is solved to propagate the electron 
  $r_{\rm t}(t,R_{\rm t}(t))$ to the next value of $t$ and of $R_{\rm t}$,
  thus providing trajectories of the electron for each trajectory of the 
  nucleus.
  The trajectories closely follow $|\phi(r|R,t)|^2$ with $t$, 
  illustrating that these are indeed the conditional trajectories corresponding 
  to the state of the electron, given the nucleus is at a certain position, and 
  given a value for the external time $t$.
  
  The considered dynamics is adiabatic in the sense of the Born-Oppenheimer 
  approximation, which is typically interpreted as the electron following the 
  nucleus ``instantaneously'', staying in an eigenstate of $\hat{H}_{\rm S}$ 
  given by \eqref{eq:hs}.
  Thus, the probability density $|\phi(r|R,t)|^2$ is approximately the 
  probability density of the Born-Oppenheimer wavefunction, cf.\ \eqref{eq:adia}.
  The Born-Oppenheimer approximation, however, only provides an approximation to 
  $|\phi(r|R,t)|$ but not to the phase of $\phi(r|R,t)$.
  Consequently, it does not provide the electronic flux density and hence not 
  the electronic motion.\cite{barth2009,scherrer2013,schild2016}
  However, the dynamics is also adiabatic for the phase of $\phi(r|R,t)$, not 
  only for its magnitude:
  Instead of solving the Born-Oppenheimer TISE \eqref{eq:hsboa}, the CDSE 
  \eqref{eq:phi} can be solved with the nucleus being the clock but without 
  explicit reference to the external time $t$.
  As explained in \cite{schild2018}, solving such a CDSE (the term
  $\hat{U} \phi$ and a term $\pt \phi$ that appears in a time- and clock-dependent 
  treatment can be neglected) provides an excellent approximation to  $\phi(r|R,t)$.
  
  As illustrated in \cite{schild2018} with the clock-dependent continuity equation
  and as can be expected from the idea of the Born-Oppenheimer approximation, the relevant clock for the electron in the adiabatic limit is the nucleus, while 
  $t$ is only affecting the dynamics indirectly as parameter for the trajectory 
  of the nucleus.
  Thus, each of the trajectories $r_{\rm t}(t,R_{\rm t}(t))$ can be read as 
  $r_{\rm t}(R_{\rm t})$, where $R_{\rm t}$ provides a possible trajectory of 
  the clock.
  It is in this sense that the model illustrates trajectories obtained from the 
  CDSE, even though the overall dynamics within the model is generated by solving 
  a TDSE for the joint system.

  
  \section{Discussion}
  \label{sec:discussion}
  
  The CDSE \eqref{eq:phi} is the generalization of the TDSE \eqref{eq:tdse} in 
  case changes in a quantum system are compared to a clock which is treated 
  quantum-mechanically, i.e., which is described by a wavefunction.
  Such a clock does not provide a single time parameter via a well-defined 
  (classical) position and velocity.
  Notwithstanding, the CDSE (like the TDSE) represents the conditional 
  dependence of the quantum system on the configuration of the clock, hence 
  there is no uncertainty of the clock configuration in the CDSE like there 
  is no uncertainty of time in the TDSE.
  The ``quantum'' effect of the clock is that different paths trough the space 
  of the clock configurations are possible.
  
  The CDSE can be used in a way that is very similar to the TDSE.
  In this article, the CDSE is used to derive the clock-dependent analogue of 
  time-dependent quantum hydrodynamics.
  A clock-dependent Hamilton-Jacobi equation as well as a clock-dependent 
  continuity equation are obtained which are similar to their time-dependent 
  counterparts, but which contain additional terms.
  In particular, in the clock-dependent versions of the equations the spatial 
  variation of the probability density of the clock is relevant.
  
  The method of characteristics can be used to solve the clock-dependent
  Hamilton-Jacobi equation, thus providing clock-dependent quantum trajectories.
  Specifically, trajectories for the clock are obtained, which depend on an 
  arbitrary parameter $\tau$, as well as trajectories for the actual system of 
  interest, which depend on the configuration of the clock.
  The existence of these trajectories is interesting in itself, as is the 
  parameter $\tau$ which comes from the way how the differential equation is 
  solved but which works very much like a normal time parameter:
  It parametrizes the trajectory of the clock, which in turn parametrizes the 
  trajectory of the system.
  
  The model presented in section \ref{sec:example} illustrates that 
  clock-dependent quantum trajectories may be a useful computational tool.
  For example, there is the problem of computing electron dynamics in molecular 
  systems interacting with strong laser fields.
  Laser pulses of attosecond duration are available to probe such electron 
  dynamics, which lead to the development of novel theoretical tools to simulate 
  the complicated laser-induced dynamics of molecules on this time scale.\cite{palacios2019}
  Especially for larger molecules, those rely on a clamped-nuclei approximation 
  of on trajectory simulations for the nuclear wavefunction.\cite{robb2018,mai2018,agostini2018book,mignolet2019,penfold2019,palacios2019}
  Additionally, many effects occurring in strong laser fields can be described 
  qualitatively with the help of simple trajectory calculations for the electrons. \cite{lewenstein1994,salieres2001,takemoto2011}
  The clock-dependent quantum trajectory formalism may provide the starting 
  point for developing simulations methods that treat both nuclei and electrons 
  in a trajectory-based picture, thus allowing to include necessary quantum 
  effects on all levels.
  A branching of the nuclear wavepacket, for example, can be described with 
  a set of trajectories for the nuclei (the clock) that give rise to a set of 
  trajectories for the electrons.
  For this purpose, next to \eqref{eq:pt1} and \eqref{eq:pt2} also equations 
  for the momenta need to be solved, which can be expected to pose similar 
  challenges like time-dependent quantum trajectories \cite{wyatt2005}.
  The discussion of these equations is, however, beyond the scope of this paper.
  Nevertheless, we note that for the considered model, \eqref{eq:pt2} can be 
  simplified by neglecting the momentum field associated with the electronic 
  wavefunction $\phi$, thus simplifying a possible propagation of the trajectories 
  without previous knowledge of the wavefunctions.
  
  Notwithstanding these possible practical applications of the formalism, 
  there are conceptual challenges.
  Next to the obvious limitation that this article is only concerned with 
  non-relativistic quantum mechanics, a central open question is:
  Where does the dynamics come from?
  According to the formalism presented here, the differential equation for the 
  trajectory of the clock \eqref{eq:pt2} shows that the momentum field 
  $\vec{P}_{\rm t}[\psi]$ generates changes of the clock configuration.
  If the wavefunction $\psi$ of the super-system, composed of the clock and the 
  quantum system of interest, is generated by solving a TDSE with some initial 
  conditions, there is a dynamics due to the ``passing'' of an unspecified 
  external time parameter.
  This is the case for the presented proton-coupled electron transfer model, 
  where an external time parameter was assumed.
  In contrast, for a truly closed system without reference to anything
  external, we expect $\psi$ to be an eigenstate of the Hamiltonian and thus 
  been given by a TISE.
  Then, $\vec{P}_{\rm t}[\psi]$ is zero if $\psi$ is real-valued.
  For $\vec{P}_{\rm t}[\psi]$ to be non-zero, and thus for a dynamics to happen 
  at all, $\psi$ needs e.g.\ to correspond to some rotating state, suitable 
  boundary conditions need to be imposed, or other modifications to the 
  formalism need to be made.
  It is thus not trivial to invent a model for purely clock-dependent quantum 
  mechanics, without reference to an external time, that shows a dynamics.
  
  The question about the ``origin of dynamics'' may be related to the ignorance of 
  relativistic effects, to the missing mechanism for deciding which path the 
  clock actually takes (i.e., the measurement or how internal observers can be 
  described), to the assumption 
  of an absolute space (time is defined relative to a clock, but space was 
  assumed to be absolute in this article), or to something completely different.
  It will be interesting to see if and how this question can be solved. 

  
  {\bf Acknowledgment}
  
  This research is supported by an Ambizione grant of the Swiss National Science 
  Foundation.
  
  \bibliography{bib}{}
  \bibliographystyle{unsrt}

\end{document}